\documentclass[a4paper,aps,prd,10pt,preprintnumbers,showpacs,twocolumn,superscriptaddress,nofootinbib,floatfix,amsmath,amssymb]{revtex4-1}
\usepackage{orcidlink}
\usepackage{graphicx,xcolor}
\usepackage{cmap}
\usepackage[utf8]{inputenc}
\usepackage[T1]{fontenc}

\def\imo{i}
\def\re#1{Re(#1)}
\def\im#1{Im(#1)}
\def\K{{\cal K}}

\begin{document}
\title{Quasi-resonances in the vicinity of Einstein-Maxwell-dilaton black hole}
\author{S. V. Bolokhov  \orcidlink{0000-0002-9533-530X}
}
\email{bolokhov-sv@rudn.ru}
\affiliation{RUDN University, 6 Miklukho-Maklaya St, Moscow, 117198, Russian Federation}
\begin{abstract}
We study massive scalar quasinormal spectra of charged Einstein--Maxwell--dilaton black holes by combining high-order WKB--Pad\'e calculations with time-domain evolution. The two approaches show close agreement in the regime where both methods are reliable, allowing controlled tracking of spectral trends across different charges and dilaton couplings. We find that increasing scalar-field mass can strongly suppress damping for several branches, signaling an approach to quasi-resonant, very long-lived oscillations. Although WKB is not expected to determine modes extremely close to the real-frequency axis with high precision, the onset of this regime is clear and appears for multiple dilaton couplings, with additional near-resonant behavior in lower-multipole sectors. The dilaton-induced shifts are substantially larger than the estimated numerical uncertainty, indicating that quasi-resonances are a robust physical signature relevant for ringdown spectroscopy in scalar-extended gravity.
\end{abstract}
\maketitle
\section{Introduction}

Quasinormal modes (QNMs) are the characteristic damped oscillations of perturbed compact objects and provide direct information about the underlying spacetime geometry \cite{Kokkotas:1999bd,Konoplya:2011qq,Bolokhov:2025rng}. In black-hole physics they are central to ringdown spectroscopy, where observed frequencies and damping rates can be used to infer mass, spin, and additional charges or couplings \cite{LIGOScientific:2016aoc,LIGOScientific:2017vwq,LIGOScientific:2020zkf,KAGRA:2021vkt,KAGRA:2013rdx}. Because QNMs are determined by the background and boundary conditions, they also offer a sensitive probe of deviations from general relativity and of extra fields predicted by extended gravity models. Understanding how these modes change when new physical scales are introduced is therefore essential for both theoretical consistency tests and future precision gravitational-wave measurements.

Dilatonic black holes are particularly important in this context because they emerge naturally in Einstein-Maxwell-dilaton and low-energy string-inspired models, and they interpolate between the Reissner-Nordstr\"{o}m limit and genuinely dilatonic charged geometries through the coupling parameter $a$. This makes them a useful laboratory for isolating how an additional scalar sector affects black-hole oscillation spectra. Although massless perturbations of dilatonic black holes have been studied in a number of works \cite{Konoplya:2001ji,Konoplya:2002ky,Zinhailo:2019rwd,Kokkotas:2017ymc,Kokkotas:2015uma,Fernando:2003wc,Fernando:2016ftj,Zinhailo:2019rwd}, a systematic analysis of massive scalar perturbations over the parameter space is still missing.

Massive scalar perturbations around compact objects provide a natural extension of the standard quasinormal-mode problem and have been investigated in many settings \cite{Konoplya:2004wg,Konoplya:2018qov,Ohashi:2004wr,Zhang:2018jgj, Bolokhov:2023ruj, Bolokhov:2024ixe, Bolokhov:2024bke, Skvortsova:2024eqi, Skvortsova:2025cah,Lutfuoglu:2026gis, Lutfuoglu:2026xlo, Lutfuoglu:2026fpx, Lutfuoglu:2025kqp, Lutfuoglu:2025eik, Lutfuoglu:2025qkt,Malik:2025czt, Malik:2025qnr, Malik:2025ava, Dubinsky:2025wns, Dubinsky:2025bvf, Lutfuoglu:2026uzy}. In contrast to the massless case, introducing $\mu$ adds an intrinsic scale to the wave dynamics, which modifies both the ringdown spectrum and the late-time signal. This makes massive fields useful not only for phenomenology, but also for testing how robust black-hole spectroscopy is when additional matter scales are present.

Several mechanisms can generate such an effective mass in practice. In braneworld-inspired scenarios, for example, bulk effects induce massive terms in the four-dimensional perturbation equations \cite{Seahra:2004fg}. A nonzero $\mu$ can also support very long-lived oscillations near specific parameter values \cite{Ohashi:2004wr,Konoplya:2004wg}. This behavior has been reported for different black-hole geometries \cite{Zhidenko:2006rs,Zinhailo:2019rwd,Churilova:2020bql,Bolokhov:2023bwm} and for other compact objects, including wormholes \cite{Churilova:2019qph}. At late times, the same mass term changes the asymptotic decay law from the familiar power-like behavior to oscillatory tails \cite{Jing:2004zb,Koyama:2001qw,Moderski:2001tk,Rogatko:2007zz,Koyama:2001ee,Koyama:2000hj,Gibbons:2008gg,Gibbons:2008rs,Dubinsky:2024jqi}. In addition, even a massless scalar can acquire an effective mass in magnetized black-hole environments \cite{Konoplya:2007yy,Konoplya:2008hj,Wu:2015fwa}. After all, the massive or effective massive fields may contribute to the very long gravitational waves observed via the Pulsar Timing Array experiment \cite{Konoplya:2023fmh}.

At the same time, long-lived modes are not universal: depending on the background and field parameters, they may be absent \cite{Zinhailo:2024jzt}. This motivates a systematic scan of the parameter space and a direct comparison between semi-analytic and time-domain approaches, which is the strategy adopted in this work.

The paper is organized as follows. In Sec.~\ref{sec:wavelike} we summarize the Einstein--Maxwell--dilaton background and the wave equation for a massive scalar perturbation. In Sec.~\ref{sec:Methods} we briefly describe the time-domain integration scheme and the high-order WKB--Pad\'e approach. The quasinormal spectra, accuracy assessment, and evidence for quasi-resonant long-lived modes are presented in Sec.~IV, and the main outcomes are summarized in the Conclusions.

\section{Basic equations}\label{sec:wavelike}

Einstein-Maxwell-dilaton (EMD) theory extends general relativity by a dilaton field nonminimally coupled to the electromagnetic sector through the coupling parameter $a$. This framework admits asymptotically flat charged black-hole solutions that interpolate between the Reissner-Nordstr\"{o}m case ($a=0$) and the string-inspired case ($a=1$). The value $a=\sqrt{3}$ has a special interpretation as the Kaluza-Klein coupling, arising from dimensional reduction of five-dimensional vacuum gravity to four dimensions.

We consider Einstein-Maxwell-dilaton theory with action
\begin{equation}\label{action}
S=\int d^4x\sqrt{-g}\left(R-2(\nabla\varphi)^2+e^{-2a\varphi}F^2\right),
\end{equation}
and a static spherically symmetric charged dilaton black-hole solution
\begin{equation}\label{metric-emd}
ds^2=\lambda^2dt^2-\lambda^{-2}dr^2-R^2d\theta^2-R^2\sin^2\theta\,d\phi^2,
\end{equation}
where
\begin{equation}\label{lambda-R}
\lambda^2=\left(1-\frac{r_+}{r}\right)\left(1-\frac{r_-}{r}\right)^{\frac{1-a^2}{1+a^2}}, \qquad
R^2=r^2\left(1-\frac{r_-}{r}\right)^{\frac{2a^2}{1+a^2}},
\end{equation}
and
\begin{equation}\label{M-Q}
2M=r_+ + \left(\frac{1-a^2}{1+a^2}\right)r_-, \qquad Q^2=\frac{r_-r_+}{1+a^2}.
\end{equation}
The extremal limit corresponds to a degenerate horizon, $r_+=r_-\equiv r_{\mathrm{ext}}$. Using Eqs.~(\ref{M-Q}), one finds
\begin{equation}
Q_{\mathrm{ext}}^2=M^2(1+a^2), \qquad Q_{\mathrm{ext}}=M\sqrt{1+a^2}.
\end{equation}
Therefore, for our normalization $M=1$ the maximal charge is
$Q_{\mathrm{ext}}=1$ for $a=0$, $Q_{\mathrm{ext}}=\sqrt{2}$ for $a=1$, and $Q_{\mathrm{ext}}=2$ for $a=\sqrt{3}$.
The dilaton and electromagnetic fields are given by
\begin{equation}\label{dilaton-EM}
e^{2a\varphi}=\left(1-\frac{r_-}{r}\right)^{\frac{2a^2}{1+a^2}}, \qquad F_{tr}=\frac{e^{2a\varphi}Q}{R^2}.
\end{equation}

For perturbation equations below, we define
\begin{equation}\label{metric}
f(r)\equiv\lambda^2(r).
\end{equation}

The covariant Klein-Gordon equation for a massive scalar field $\Phi$ is
\begin{equation}\label{KGg}
\frac{1}{\sqrt{-g}}\partial_\mu \left(\sqrt{-g}g^{\mu \nu}\partial_\nu\Phi\right)-\mu^2\Phi=0.
\end{equation}
After separation of variables in the background metric (\ref{metric-emd}), with $f(r)$ defined by Eq.~(\ref{metric}), Eq.~(\ref{KGg}) takes the Schrödinger wavelike form \cite{Carter:1968ks,Konoplya:2018arm}:
\begin{equation}\label{wave-equation}
\dfrac{d^2 \Psi}{dr_*^2}+(\omega^2-V(r))\Psi=0,
\end{equation}
where the ``tortoise coordinate'' $r_*$ is defined as follows:
\begin{equation}\label{tortoise}
dr_*\equiv\frac{dr}{f(r)}.
\end{equation}

The effective potential for the massive scalar field has the form
\begin{equation}\label{potentialScalar}
V(r)=f(r)\left(\mu^2+\frac{\ell(\ell+1)}{R(r)^2}\right)+\frac{1}{R(r)}\cdot\frac{d^2 R(r)}{dr_*^2},
\end{equation}
where $\ell=0,1,2,\ldots$ are the multipole numbers.

\begin{figure}
\resizebox{\linewidth}{!}{\includegraphics{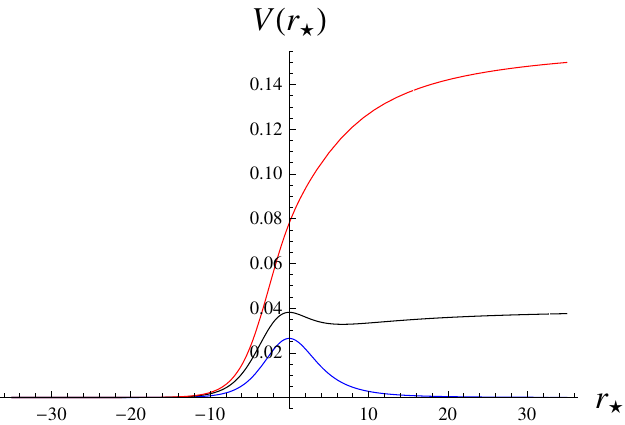}}
\caption{Effective potential $V(r_*)$ for massive scalar perturbations with $\ell=0$ and $a=0$ ($M=1$): $\mu=0$ (blue), $\mu=0.2$ (black), and $\mu=0.4$ (red).}\label{fig:pot1}
\end{figure}

\begin{figure}
\resizebox{\linewidth}{!}{\includegraphics{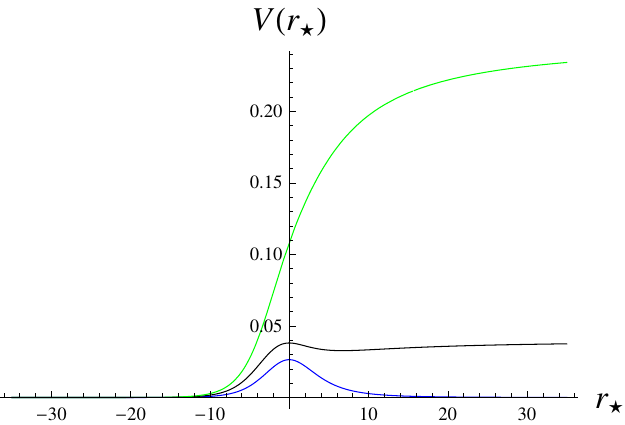}}
\caption{Effective potential $V(r_*)$ for massive scalar perturbations with $\ell=0$ and $a=1$ ($M=1$): $\mu=0$ (blue), $\mu=0.2$ (black), and $\mu=0.5$ (green).}\label{fig:pot2}
\end{figure}

Figures~\ref{fig:pot1} and \ref{fig:pot2} show typical effective potentials for the massive scalar field. In both cases, the potential has the standard barrier shape required for quasinormal-mode boundary-value problems; increasing $\mu$ raises the potential at large $r_*$ (toward the asymptotic value set by the mass term) and modifies the barrier profile near its maximum. A larger dilaton coupling $a$ also changes the barrier quantitatively, which leads to noticeable shifts of the quasinormal spectrum discussed in the next sections.

\section{Time-domain integration and WKB method}\label{sec:Methods}
We first compute the signal in the time domain and use it as an independent benchmark for the frequency-domain results. Writing the wave equation in light-cone coordinates $u=t-r_*$ and $v=t+r_*$, we evolve the profile with the Gundlach-Price-Pullin finite-difference scheme \cite{Gundlach:1993tp}
\begin{eqnarray}
\Psi\left(N\right)&=&\Psi\left(W\right)+\Psi\left(E\right)-\Psi\left(S\right)\nonumber\\
&&- \Delta^2V\left(S\right)\frac{\Psi\left(W\right)+\Psi\left(E\right)}{8}+{\cal O}\left(\Delta^4\right),\label{Discretization}
\end{eqnarray}
where $N\equiv(u+\Delta,v+\Delta)$, $W\equiv(u+\Delta,v)$, $E\equiv(u,v+\Delta)$, and $S\equiv(u,v)$. The quasinormal frequency is extracted from the ringdown interval, while late-time tails are used as an additional consistency check. This integration approach has been extensively tested in black-hole perturbation problems \cite{Dubinsky:2025fwv,Bolokhov:2024ixe,Malik:2024nhy,Konoplya:2024hfg,Bolokhov:2023bwm,Skvortsova:2024atk,Malik:2023bxc,Varghese:2011ku,Skvortsova:2023zmj,Skvortsova:2024wly,Lutfuoglu:2025hjy,Bolokhov:2023dxq,Churilova:2021tgn,Dubinsky:2024nzo,Abdalla:2012si,Bolokhov:2026eqf, Arbelaez:2026eaz, Arbelaez:2025gwj}.

\begin{figure}
\resizebox{\linewidth}{!}{\includegraphics{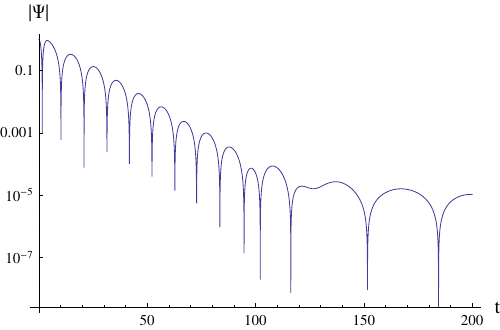}}
\caption{Time-domain profile for $\ell=1$, $a=0$, $M=1$, $Q=0.3$, and $\mu=0.1$. The Prony fit of the ringdown yields the fundamental mode $\omega_{\mathrm{Prony}}=0.301956-0.0954401\,i$. For the same parameters, the WKB16 value in Table~II is $\omega_{\mathrm{WKB}}=0.301918-0.095489\,i$. The relative discrepancy is about $0.02\%$ in the complex-frequency norm (with separate differences $0.013\%$ in $\re{\omega}$ and $0.051\%$ in $|\im{\omega}|$), which may be viewed as the relative WKB error for this mode.}
\label{fig:TD1}
\end{figure}

\begin{figure}
\resizebox{\linewidth}{!}{\includegraphics{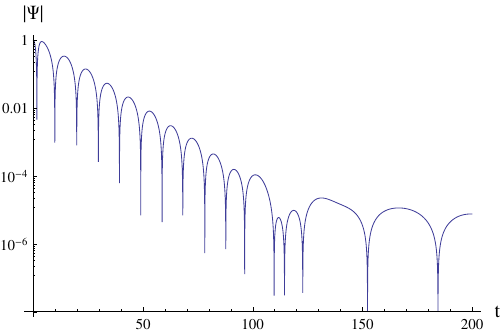}}
\caption{Time-domain profile for $\ell=1$, $a=1$, $M=1$, $Q=0.7$, and $\mu=0.1$. The Prony fit gives $\omega_{\mathrm{Prony}}=0.324979-0.0980759\,i$. For the same parameters, the WKB16 value in Table~IV is $\omega_{\mathrm{WKB}}=0.324963-0.098140\,i$. The relative discrepancy is about $0.02\%$ in the complex-frequency norm (with separate differences $0.005\%$ in $\re{\omega}$ and $0.065\%$ in $|\im{\omega}|$), which is consistent with the expected WKB error for the fundamental mode in this sector.}
\label{fig:TD2}
\end{figure}

\begin{table}
\begin{tabular}{c c c c c}
\hline
Q & $\mu$ & WKB16 ($\tilde{m}=8$) & WKB14 ($\tilde{m}=7$) & diff.  \\
\hline
$0$ & $0$ & $0.110473-0.104954 i$ & $0.110394-0.104612 i$ & $0.230\%$\\
$0$ & $0.1$ & $0.112880-0.096990 i$ & $0.115824-0.095846 i$ & $2.12\%$\\
$0$ & $0.2$ & $0.125466-0.067499 i$ & $0.114439-0.074719 i$ & $9.25\%$\\
$0.3$ & $0$ & $0.112280-0.105316 i$ & $0.112187-0.104974 i$ & $0.230\%$\\
$0.3$ & $0.1$ & $0.114688-0.097487 i$ & $0.117507-0.095715 i$ & $2.21\%$\\
$0.3$ & $0.25$ & $0.202657-0.005137 i$ & $0.204604-0.003927 i$ & $1.13\%$\\
$0.7$ & $0$ & $0.122065-0.105906 i$ & $0.121896-0.105537 i$ & $0.251\%$\\
$0.7$ & $0.05$ & $0.122697-0.104151 i$ & $0.122482-0.103718 i$ & $0.300\%$\\
$0.7$ & $0.1$ & $0.124550-0.098864 i$ & $0.124427-0.096704 i$ & $1.36\%$\\
$0.7$ & $0.15$ & $0.127087-0.090783 i$ & $0.127680-0.090443 i$ & $0.437\%$\\
$0.7$ & $0.2$ & $0.126740-0.082169 i$ & $0.123007-0.085888 i$ & $3.49\%$\\
$0.9$ & $0$ & $0.132269-0.102190 i$ & $0.132290-0.102140 i$ & $0.0325\%$\\
$0.9$ & $0.05$ & $0.132946-0.100697 i$ & $0.132970-0.100639 i$ & $0.0376\%$\\
$0.9$ & $0.1$ & $0.134170-0.095758 i$ & $0.134532-0.095820 i$ & $0.222\%$\\
$0.9$ & $0.15$ & $0.138201-0.088979 i$ & $0.138123-0.089009 i$ & $0.0513\%$\\
$0.9$ & $0.2$ & $0.141514-0.077739 i$ & $0.140133-0.077766 i$ & $0.856\%$\\
\hline
\end{tabular}
\caption{QNMs of the $\ell=0$ scalar perturbations of the dilaton black hole ($M=1$, $a=0$), calculated using the WKB formula at different orders and Pad\'e approximants. From here on, $\tilde{m}$ is defined as in \cite{Konoplya:2019hlu}. The deviation is given in percent.}
\end{table}
\begin{table}
\begin{tabular}{c c c c c}
\hline
Q & $\mu$ & WKB16 ($\tilde{m}=8$) & WKB14 ($\tilde{m}=7$) & diff.  \\
\hline
$0$ & $0$ & $0.292936-0.097660 i$ & $0.292936-0.097660 i$ & $0\%$\\
$0$ & $0.1$ & $0.297416-0.094957 i$ & $0.297416-0.094957 i$ & $0\%$\\
$0$ & $0.2$ & $0.310957-0.086593 i$ & $0.310957-0.086593 i$ & $0\%$\\
$0$ & $0.25$ & $0.321199-0.080040 i$ & $0.321199-0.080040 i$ & $0.\times 10^{\text{-4}}\%$\\
$0$ & $0.3$ & $0.333777-0.071658 i$ & $0.333778-0.071658 i$ & $0.\times 10^{\text{-4}}\%$\\
$0$ & $0.35$ & $0.348640-0.061174 i$ & $0.348641-0.061174 i$ & $0.00043\%$\\
$0$ & $0.4$ & $0.365599-0.048285 i$ & $0.365587-0.048280 i$ & $0.00343\%$\\
$0$ & $0.45$ & $0.384784-0.035853 i$ & $0.386409-0.036699 i$ & $0.474\%$\\
$0.3$ & $0$ & $0.297537-0.098106 i$ & $0.297537-0.098106 i$ & $0\%$\\
$0.3$ & $0.1$ & $0.301918-0.095489 i$ & $0.301918-0.095489 i$ & $0\%$\\ 
$0.3$ & $0.2$ & $0.315155-0.087395 i$ & $0.315155-0.087395 i$ & $0\%$\\
$0.3$ & $0.25$ & $0.325163-0.081054 i$ & $0.325163-0.081054 i$ & $0\%$\\
$0.3$ & $0.3$ & $0.337454-0.072949 i$ & $0.337454-0.072949 i$ & $0.\times 10^{\text{-4}}\%$\\
$0.3$ & $0.35$ & $0.351979-0.062817 i$ & $0.351981-0.062818 i$ & $0.0007\%$\\
$0.3$ & $0.4$ & $0.368554-0.050364 i$ & $0.368620-0.050370 i$ & $0.0177\%$\\
$0.3$ & $0.45$ & $0.386220-0.036167 i$ & $0.386371-0.036939 i$ & $0.203\%$\\
$0.7$ & $0$ & $0.322770-0.099350 i$ & $0.322770-0.099350 i$ & $0\%$\\
$0.7$ & $0.05$ & $0.323738-0.098812 i$ & $0.323738-0.098812 i$ & $0\%$\\
$0.7$ & $0.1$ & $0.326647-0.097184 i$ & $0.326647-0.097184 i$ & $0\%$\\
$0.7$ & $0.15$ & $0.331511-0.094430 i$ & $0.331511-0.094430 i$ & $0\%$\\
$0.7$ & $0.2$ & $0.338350-0.090482 i$ & $0.338350-0.090482 i$ & $0\%$\\
$0.7$ & $0.25$ & $0.347189-0.085236 i$ & $0.347189-0.085236 i$ & $0\%$\\
$0.7$ & $0.3$ & $0.358039-0.078536 i$ & $0.358039-0.078536 i$ & $0.\times 10^{\text{-4}}\%$\\
$0.7$ & $0.35$ & $0.370873-0.070174 i$ & $0.370873-0.070174 i$ & $0.\times 10^{\text{-4}}\%$\\
$0.7$ & $0.4$ & $0.385584-0.059900 i$ & $0.385577-0.059888 i$ & $0.00345\%$\\
$0.7$ & $0.45$ & $0.401880-0.047482 i$ & $0.401892-0.047595 i$ & $0.0281\%$\\
$0.7$ & $0.5$ & $0.446093-0.012075 i$ & $0.447531-0.009921 i$ & $0.580\%$\\
$0.9$ & $0$ & $0.352580-0.097191 i$ & $0.352580-0.097191 i$ & $0\%$\\
$0.9$ & $0.05$ & $0.353417-0.096782 i$ & $0.353417-0.096782 i$ & $0\%$\\
$0.9$ & $0.1$ & $0.355934-0.095546 i$ & $0.355934-0.095546 i$ & $0\%$\\
$0.9$ & $0.15$ & $0.360141-0.093449 i$ & $0.360141-0.093449 i$ & $0.\times 10^{\text{-4}}\%$\\
$0.9$ & $0.2$ & $0.366056-0.090436 i$ & $0.366056-0.090436 i$ & $0.\times 10^{\text{-4}}\%$\\
$0.9$ & $0.25$ & $0.373700-0.086418 i$ & $0.373700-0.086418 i$ & $0\%$\\
$0.9$ & $0.3$ & $0.383088-0.081270 i$ & $0.383088-0.081270 i$ & $0\%$\\
$0.9$ & $0.35$ & $0.394212-0.074818 i$ & $0.394212-0.074817 i$ & $0.00011\%$\\
$0.9$ & $0.4$ & $0.407007-0.066842 i$ & $0.407008-0.066843 i$ & $0.00026\%$\\
$0.9$ & $0.45$ & $0.421320-0.057124 i$ & $0.421321-0.057134 i$ & $0.00232\%$\\
$0.9$ & $0.5$ & $0.437138-0.045249 i$ & $0.436788-0.045207 i$ & $0.0801\%$\\
\hline
\end{tabular}
\caption{QNMs of the $\ell=1$ scalar perturbations of the dilaton black hole ($M=1$, $a=0$), calculated using the WKB formula at different orders and Pad\'e approximants. The deviation is given in percent.}
\end{table}

For semi-analytic estimates we then apply the higher-order WKB expansion. Quasinormal modes satisfy outgoing/ingoing boundary conditions,
\begin{equation}\label{boundaryconditions}
\Psi(r_*\to\pm\infty)\propto e^{\pm\imo \omega r_*},
\end{equation}
and for a single dominant peak the WKB formula gives
\begin{eqnarray}\label{WKBformula-spherical}
\omega^2&=&V_0+A_2(\K^2)+A_4(\K^2)+A_6(\K^2)+\ldots\\\nonumber&-&\imo \K\sqrt{-2V_2}\left(1+A_3(\K^2)+A_5(\K^2)+A_7(\K^2)+\ldots\right),
\end{eqnarray}
with
\begin{equation}
\K=n+\frac{1}{2}, \quad n=0,1,2,\ldots .
\end{equation}
Here $V_0$ and $V_2$ are, respectively, the value of the effective potential and its second tortoise derivative at the maximum. Higher-order terms $A_i$ depend on derivatives up to order $2i$; explicit expressions are available in \cite{Iyer:1986np,Konoplya:2003ii,Matyjasek:2017psv}. Here, following \cite{Konoplya:2026fqh}, we used 14th and 16th WKB orders, which is known to produce sufficoently accurate results for massive fields.

A key point for massive fields is that the effective potential is not always of the simplest barrier type: depending on $\mu$ and the black-hole parameters, it may develop additional turning points or even lose its local maximum. If no maximum exists, the standard WKB construction is not applicable. When the potential still has a well-defined dominant maximum (even in the presence of extra turning points), the WKB approximation usually remains accurate for low overtones; in such cases we additionally verify the result against time-domain evolution. In the tables below we quote high-order WKB values with Pad\'e improvement and compare neighboring approximants to monitor convergence \cite{Konoplya:2019hlu,Abdalla:2005hu,Fernando:2016ftj,Bolokhov:2025egl,Konoplya:2005sy,Konoplya:2001ji,Konoplya:2006ar,Kokkotas:2010zd,Churilova:2019qph,Malik:2025erb,Pathrikar:2025gzu,Momennia:2018hsm,Guo:2020caw,Konoplya:2009hv,Eniceicu:2019npi,Gonzalez:2022ote, Bolokhov:2025aqy, Bolokhov:2025lnt, Skvortsova:2024msa, Malik:2026lfj}.
\begin{table}
\begin{tabular}{c c c c c}
\hline
Q & $\mu$ & WKB16 ($\tilde{m}=8$) & WKB14 ($\tilde{m}=7$) & diff.  \\
\hline
$0$ & $0$ & $0.110473-0.104954 i$ & $0.110394-0.104612 i$ & $0.230\%$\\
$0$ & $0.1$ & $0.112880-0.096990 i$ & $0.115824-0.095846 i$ & $2.12\%$\\
$0$ & $0.2$ & $0.125466-0.067499 i$ & $0.114439-0.074719 i$ & $9.25\%$\\
$0.3$ & $0$ & $0.112264-0.105342 i$ & $0.112172-0.105000 i$ & $0.230\%$\\
$0.3$ & $0.1$ & $0.114668-0.097511 i$ & $0.117505-0.095748 i$ & $2.22\%$\\
$0.3$ & $0.25$ & $0.202523-0.005226 i$ & $0.204484-0.003996 i$ & $1.14\%$\\
$0.7$ & $0$ & $0.121422-0.107173 i$ & $0.121272-0.106829 i$ & $0.232\%$\\
$0.7$ & $0.05$ & $0.122014-0.105397 i$ & $0.121817-0.104988 i$ & $0.281\%$\\
$0.7$ & $0.1$ & $0.123802-0.100064 i$ & $0.123654-0.097630 i$ & $1.53\%$\\
$0.7$ & $0.15$ & $0.126149-0.091818 i$ & $0.126841-0.091652 i$ & $0.456\%$\\
$0.7$ & $0.2$ & $0.124809-0.085115 i$ & $0.126019-0.103870 i$ & $12.4\%$\\
$1.4$ & $0$ & $0.235235-0.091412 i$ & $0.234356-0.091741 i$ & $0.372\%$\\
$1.4$ & $0.05$ & $0.235287-0.091019 i$ & $0.234358-0.091339 i$ & $0.389\%$\\
$1.4$ & $0.1$ & $0.235681-0.089679 i$ & $0.233860-0.090146 i$ & $0.746\%$\\
$1.4$ & $0.15$ & $0.236920-0.085853 i$ & $0.236573-0.087122 i$ & $0.522\%$\\
$1.4$ & $0.2$ & $0.234964-0.089428 i$ & $0.229025-0.088519 i$ & $2.39\%$\\
\hline
\end{tabular}
\caption{QNMs of the $\ell=0$ scalar perturbations of the dilaton black hole ($M=1$, $a=1$), calculated using the WKB formula at different orders and Pad\'e approximants. The deviation is given in percent.}
\end{table}
\begin{table}
\begin{tabular}{c c c c c}
\hline
Q & $\mu$ & WKB16 ($\tilde{m}=8$) & WKB14 ($\tilde{m}=7$) & diff. \\
\hline
$0$ & $0$ & $0.292936-0.097660 i$ & $0.292936-0.097660 i$ & $0\%$\\
$0$ & $0.1$ & $0.297416-0.094957 i$ & $0.297416-0.094957 i$ & $0\%$\\
$0$ & $0.2$ & $0.310957-0.086593 i$ & $0.310957-0.086593 i$ & $0\%$\\
$0$ & $0.25$ & $0.321199-0.080040 i$ & $0.321199-0.080040 i$ & $0.\times 10^{\text{-4}}\%$\\
$0$ & $0.3$ & $0.333777-0.071658 i$ & $0.333778-0.071658 i$ & $0.\times 10^{\text{-4}}\%$\\
$0$ & $0.35$ & $0.348640-0.061174 i$ & $0.348641-0.061174 i$ & $0.00043\%$\\
$0$ & $0.4$ & $0.365599-0.048285 i$ & $0.365587-0.048280 i$ & $0.00343\%$\\
$0$ & $0.45$ & $0.384784-0.035853 i$ & $0.386409-0.036699 i$ & $0.474\%$\\
$0.3$ & $0$ & $0.297500-0.098127 i$ & $0.297500-0.098127 i$ & $0\%$\\
$0.3$ & $0.1$ & $0.301881-0.095509 i$ & $0.301881-0.095509 i$ & $0\%$\\
$0.3$ & $0.2$ & $0.315120-0.087409 i$ & $0.315120-0.087409 i$ & $0\%$\\
$0.3$ & $0.25$ & $0.325130-0.081064 i$ & $0.325130-0.081064 i$ & $0\%$\\
$0.3$ & $0.3$ & $0.337422-0.072954 i$ & $0.337422-0.072954 i$ & $0.\times 10^{\text{-4}}\%$\\
$0.3$ & $0.35$ & $0.351949-0.062817 i$ & $0.351951-0.062818 i$ & $0.0007\%$\\
$0.3$ & $0.4$ & $0.368527-0.050358 i$ & $0.368596-0.050366 i$ & $0.0187\%$\\
$0.3$ & $0.45$ & $0.386194-0.036160 i$ & $0.386349-0.036937 i$ & $0.204\%$\\
$0.7$ & $0$ & $0.321059-0.100372 i$ & $0.321059-0.100372 i$ & $0\%$\\
$0.7$ & $0.1$ & $0.324963-0.098140 i$ & $0.324963-0.098140 i$ & $0\%$\\
$0.7$ & $0.2$ & $0.336749-0.091239 i$ & $0.336749-0.091239 i$ & $0\%$\\
$0.7$ & $0.3$ & $0.356580-0.078964 i$ & $0.356580-0.078964 i$ & $0\%$\\
$0.7$ & $0.4$ & $0.384340-0.059863 i$ & $0.384322-0.059863 i$ & $0.00463\%$\\
$0.7$ & $0.5$ & $0.451908-0.004618 i$ & $0.453133-0.003601 i$ & $0.352\%$\\
$1.4$ & $0$ & $0.631929-0.088821 i$ & $0.631930-0.088821 i$ & $0.00003\%$\\
$1.4$ & $0.1$ & $0.632530-0.088486 i$ & $0.632530-0.088486 i$ & $0.00003\%$\\
$1.4$ & $0.2$ & $0.634343-0.087461 i$ & $0.634343-0.087461 i$ & $0.00003\%$\\
$1.4$ & $0.3$ & $0.637401-0.085688 i$ & $0.637401-0.085688 i$ & $0.00002\%$\\
$1.4$ & $0.4$ & $0.641754-0.083057 i$ & $0.641754-0.083057 i$ & $0.00002\%$\\
$1.4$ & $0.5$ & $0.647464-0.079385 i$ & $0.647464-0.079385 i$ & $0\%$\\
\hline
\end{tabular}
\caption{QNMs of the $\ell=1$ scalar perturbations of the dilaton black hole ($M=1$, $a=1$), calculated using the WKB formula at different orders and Pad\'e approximants. The deviation is given in percent.}
\end{table}
\begin{table}
\begin{tabular}{c c c c c}
\hline
Q & $\mu$ & WKB16 ($\tilde{m}=8$) & WKB14 ($\tilde{m}=7$) & diff.  \\
\hline
$0$ & $0$ & $0.110473-0.104954 i$ & $0.110394-0.104612 i$ & $0.230\%$\\
$0$ & $0.05$ & $0.111037-0.102942 i$ & $0.110917-0.102503 i$ & $0.300\%$\\
$0$ & $0.1$ & $0.112880-0.096990 i$ & $0.115824-0.095846 i$ & $2.12\%$\\
$0$ & $0.15$ & $0.115158-0.087871 i$ & $0.115413-0.087274 i$ & $0.449\%$\\
$0$ & $0.2$ & $0.125466-0.067499 i$ & $0.114439-0.074719 i$ & $9.25\%$\\
$0.3$ & $0$ & $0.112234-0.105386 i$ & $0.112142-0.105045 i$ & $0.230\%$\\
$0.3$ & $0.05$ & $0.112806-0.103414 i$ & $0.112669-0.102983 i$ & $0.296\%$\\
$0.3$ & $0.1$ & $0.114630-0.097551 i$ & $0.117504-0.095804 i$ & $2.23\%$\\
$0.3$ & $0.15$ & $0.116948-0.088596 i$ & $0.117242-0.087989 i$ & $0.460\%$\\
$0.3$ & $0.2$ & $0.153331-0.078763 i$ & $0.117419-0.074899 i$ & $21.0\%$\\
$0.3$ & $0.25$ & $0.202297-0.005376 i$ & $0.204283-0.004114 i$ & $1.16\%$\\
$1.$ & $0$ & $0.132649-0.113640 i$ & $0.132743-0.113663 i$ & $0.0556\%$\\
$1.$ & $0.05$ & $0.133083-0.112037 i$ & $0.133184-0.112055 i$ & $0.0589\%$\\
$1.$ & $0.1$ & $0.134344-0.107435 i$ & $0.134477-0.107283 i$ & $0.118\%$\\
$1.$ & $0.15$ & $0.136540-0.099553 i$ & $0.136603-0.099529 i$ & $0.0393\%$\\
$1.$ & $0.2$ & $0.140868-0.090440 i$ & $0.138681-0.089588 i$ & $1.40\%$\\
$1.$ & $0.25$ & $0.134835-0.087701 i$ & $0.135073-0.089533 i$ & $1.15\%$\\
\hline
\end{tabular}
\caption{QNMs of the $\ell=0$ scalar perturbations of the dilaton black hole ($M=1$, $a=\sqrt{3}$), calculated using the WKB formula at different orders and Pad\'e approximants. The deviation is given in percent.}
\end{table}

\section{Quasinormal modes}

The tabulated spectra show two clearly different accuracy regimes. For the $\ell=1$ fundamental mode, the WKB16 and WKB14 values are almost identical for most points: for $a=0$ the relative difference is typically zero to the shown digits and stays below $0.58\%$ (maximal at $Q=0.7$, $\mu=0.5$), while for $a=1$ it remains below $0.352\%$ (again at $Q=0.7$, $\mu=0.5$). For $\ell=0$, the sensitivity to the approximation order is stronger, especially at larger $\mu$: the discrepancy reaches $9.25\%$ at $(a,Q,\mu)=(0,0,0.2)$, $12.4\%$ at $(1,0.7,0.2)$, and $21.0\%$ at $(\sqrt{3},0.3,0.2)$. This pattern is consistent with the fact that the low-multipole massive-field potential is more prone to changes in peak structure. In the $\ell=1$ sector, therefore, the expected relative error inferred from neighboring WKB orders is sub-percent, and the direct time-domain benchmark below gives an even smaller reference value ($\sim0.02\%$).

Tables I and II correspond to $a=0$, Tables III and IV correspond to $a=1$, and Table V corresponds to $a=\sqrt{3}$; all quoted frequencies in this section refer to the fundamental overtone $n=0$.

\begin{figure*}
\resizebox{\linewidth}{!}{\includegraphics{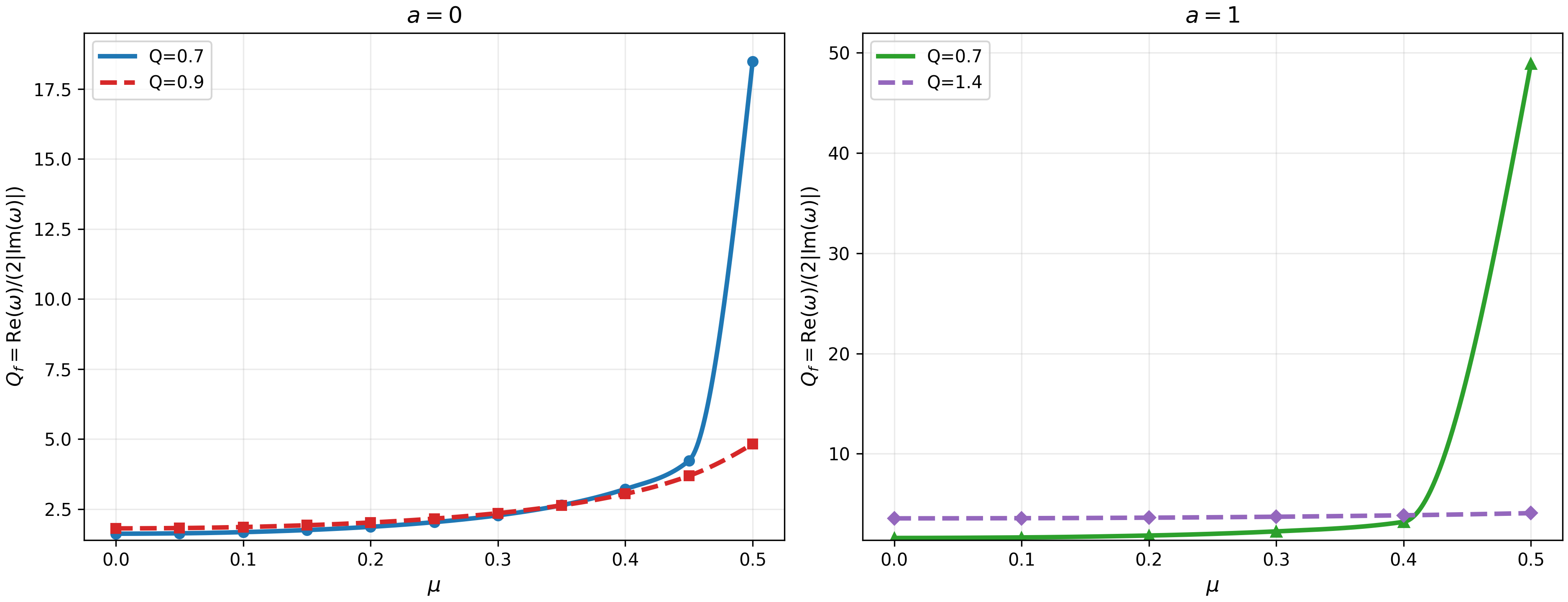}}
\caption{Quality factors $Q_f=\re{\omega}/(2|\im{\omega}|)$ for the fundamental $\ell=1$ scalar mode as functions of the field mass $\mu$, constructed from WKB16 frequencies in the updated tables. To improve visibility, only selected higher-charge branches are shown. Left panel: $a=0$ with $Q=0.7$ and $Q=0.9$. Right panel: $a=1$ with $Q=0.7$ and $Q=1.4$. Markers denote tabulated values; lines are shape-preserving cubic interpolations evaluated on a dense $\mu$ grid and used only as visual guides.}
\label{fig:qf-mu-l1}
\end{figure*}

To quantify damping for the denser and more robust $\ell=1$ dataset, we use the dimensionless quality factor
\begin{equation}
Q_f\equiv\frac{\re{\omega}}{2|\im{\omega}|},
\end{equation}
which gives the number of oscillations over one e-folding damping time. For $a=0$, $Q_f$ increases strongly with $\mu$: at $Q=0$, it grows from $1.50$ at $\mu=0$ to $5.37$ at $\mu=0.45$; at $Q=0.7$, it grows from $1.62$ to $18.47$ by $\mu=0.5$. For $Q=0.9$, the increase is more moderate (up to $Q_f=4.83$ at $\mu=0.5$). For $a=1$, the same trend holds: at $Q=0.3$, $Q_f$ rises from $1.52$ ($\mu=0$) to $5.34$ ($\mu=0.45$), while at $Q=0.7$ it reaches $48.93$ at $\mu=0.5$, corresponding to very weak damping. At the same time, the high-charge branch $Q=1.4$ has much larger oscillation frequency (from $\re{\omega}=0.631929$ to $0.647464$) but only moderate quality-factor growth ($3.56\to4.08$), because its damping remains relatively large. These trends are summarized in Fig.~\ref{fig:qf-mu-l1}.

\begin{figure}
\resizebox{\linewidth}{!}{\includegraphics{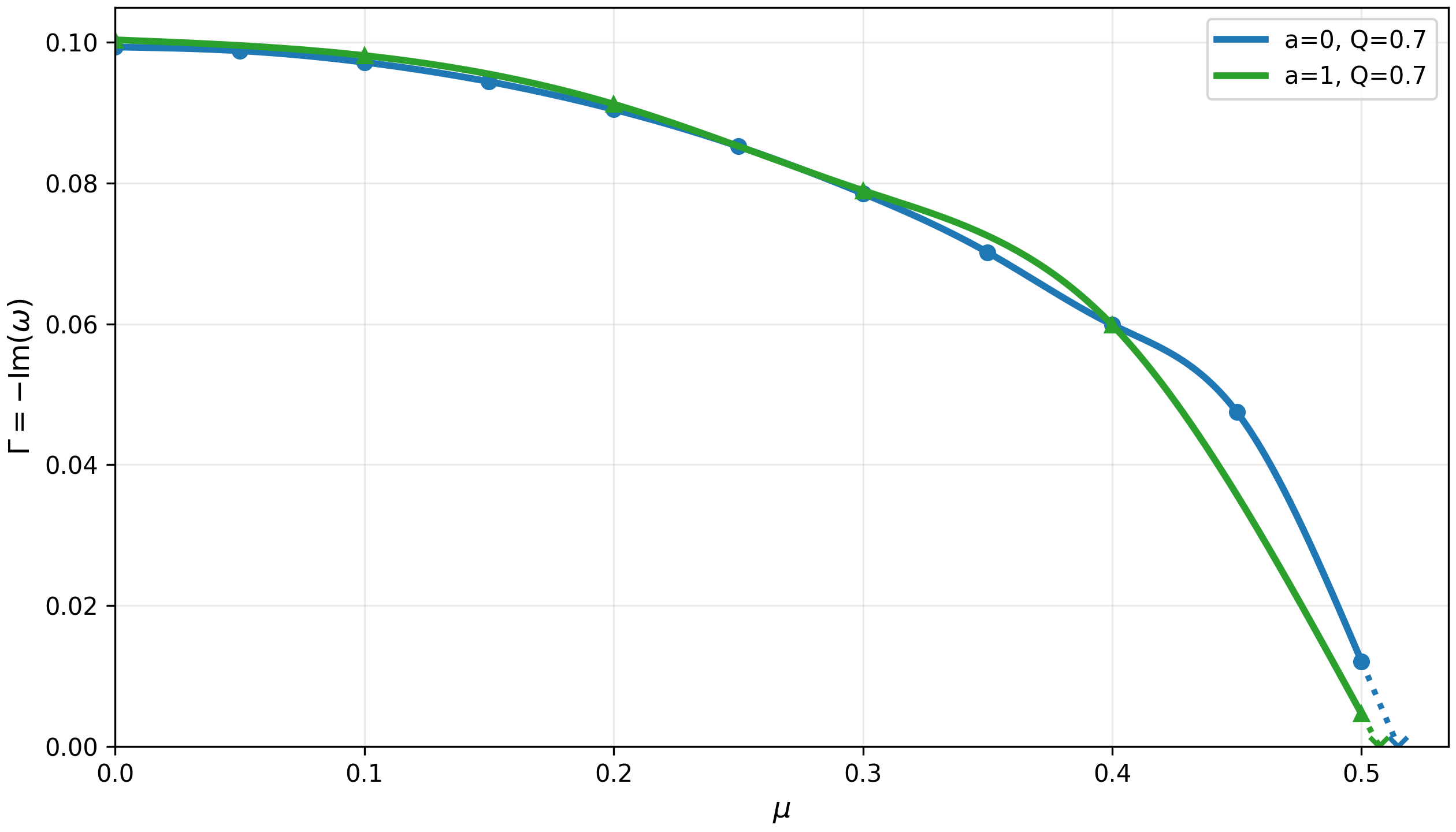}}
\caption{Damping rate $\Gamma=-\im{\omega}$ for the fundamental $\ell=1$ scalar mode as a function of the field mass $\mu$, using WKB16 frequencies from the updated tables at fixed charge $Q=0.7$. Blue circles correspond to $a=0$, and green triangles to $a=1$. Markers denote tabulated values; solid lines are smooth cubic-spline interpolations on the tabulated interval. Dotted segments show smooth extrapolation toward $\Gamma=0$, and crosses mark the corresponding formal zero crossings.}
\label{fig:gamma-mu-l1}
\end{figure}

A complementary characterization is provided by the damping rate
\begin{equation}
\Gamma\equiv-\im{\omega},
\end{equation}
shown in Fig.~\ref{fig:gamma-mu-l1} for the same $\ell=1$ modes. The decrease of $\Gamma$ with mass is explicit in the tables: for $a=0$, $Q=0$, $\Gamma$ falls from $0.097660$ at $\mu=0$ to $0.035853$ at $\mu=0.45$; for $a=0$, $Q=0.7$, it falls from $0.099350$ to $0.012075$ at $\mu=0.5$. The strongest suppression appears for $a=1$, $Q=0.7$, where $\Gamma$ drops from $0.100372$ ($\mu=0$) to $0.004618$ ($\mu=0.5$). By contrast, the $a=1$, $Q=1.4$ branch is much less sensitive to $\mu$ ($0.088821\to0.079385$), in agreement with the curve in Fig.~\ref{fig:gamma-mu-l1}. At the same $(Q,\mu)=(0.7,0.5)$, the damping rate changes from $0.012075$ for $a=0$ to $0.004618$ for $a=1$, i.e. by about $62\%$, which is much larger than the expected numerical uncertainty. The dashed extrapolations in Fig.~\ref{fig:gamma-mu-l1} therefore provide a clear indication of quasi-resonances, namely modes becoming arbitrarily long-lived when $\mu$ approaches critical values. Very close to $\Gamma=0$ (the real axis in the complex-$\omega$ plane), WKB is not sufficiently reliable to resolve the mode itself, but the extrapolated trend and the marked zero crossings show this approach unambiguously.

The $\ell=0$ tables show related but less uniform behavior. For $a=0$, the mode at $Q=0$ evolves from $0.110473-0.104954 i$ ($\mu=0$) to $0.125466-0.067499 i$ ($\mu=0.2$), and the entry at $\mu=0.25$ is purely real ($\omega=0.21651$). For $Q=0.3$, the mode at $\mu=0.25$ is already near quasi-resonant, $\omega=0.202657-0.005137 i$. For $a=1$, the same near-resonant pattern is present at $Q=0.3$, $\mu=0.25$ ($\omega=0.202523-0.005226 i$), while the larger charge $Q=1.4$ stays more damped in the shown interval ($\omega=0.235235-0.091412 i$ at $\mu=0$ and $0.234964-0.089428 i$ at $\mu=0.2$). For $a=\sqrt{3}$, the behavior is markedly different near extremality: at $Q=1.99$, the dominant complex-frequency branch is highly damped and almost $\mu$-independent in the range $0\le\mu\le0.2$ ($\omega\approx1.15775-0.831 i$).

Finally, the direct time-domain check supports the same picture in the region where the potential peak is well defined. For the representative case shown in Fig.~\ref{fig:TD1} ($\ell=1$, $a=0$, $Q=0.3$, $\mu=0.1$), the Prony value $\omega_{\mathrm{Prony}}=0.301956-0.0954401i$ and the WKB16 value $\omega_{\mathrm{WKB}}=0.301918-0.095489i$ differ by about $0.02\%$. A second example in Fig.~\ref{fig:TD2} ($\ell=1$, $a=1$, $Q=0.7$, $\mu=0.1$) shows a similarly small discrepancy: $\omega_{\mathrm{Prony}}=0.324979-0.0980759i$ versus $\omega_{\mathrm{WKB}}=0.324963-0.098140i$, again at about $0.02\%$ in the complex-frequency norm. Together with the small WKB-order differences in the $\ell=1$ tables, this agreement indicates that the quasi-resonant trends inferred from the tables and extrapolations are robust in the corresponding parameter domain.

Notice that the quasinormal modes obtained here can be used for straightforward estimations of grey-body factors of black holes by employing the correspondence between these two quantities~\cite{Konoplya:2024vuj,Konoplya:2024lir, Bolokhov:2024otn, Malik:2024cgb, Dubinsky:2024vbn}.  In particular, in the eikonal regime $\ell \gg 1$, the transmission coefficient (grey-body factor) can be expressed in terms of the fundamental quasinormal frequency $\omega_0$:
\begin{equation}
\Gamma_{\ell}(\omega) \approx \frac{1}{1 + \exp\!\left(2\pi\frac{\omega^2 - \mathrm{Re}(\omega_0)^2}{4\,\mathrm{Re}\,\omega_0\,\mathrm{Im}\,\omega_0}\right)}.
\end{equation}
Once the quasinormal spectrum is known, one can reconstruct the qualitative behavior of the transmission coefficients without solving the scattering problem explicitly. This provides an efficient alternative to direct numerical integration of the wave equation and is particularly useful for parametric studies of black hole geometries. Furthermore, inclusion of higher overtones improves the accuracy of this correspondence, especially away from the strict eikonal limit, allowing one to capture deviations from a simple single-barrier scattering picture. However, it should be emphasized that this approximation may lose accuracy for low multipole numbers or in the presence of complicated effective potentials, such as those with multiple extrema.

\section{Conclusions}

We have analyzed massive scalar quasinormal modes of Einstein--Maxwell--dilaton black holes for $a=0$, $a=1$, and $a=\sqrt{3}$, combining high-order WKB (with Pad\'e improvement) and time-domain evolution. For the dominant $\ell=1$ sector, the WKB convergence is typically sub-percent, and the direct benchmark at $(\ell,a,Q,\mu)=(1,0,0.3,0.1)$ gives only about $0.02\%$ difference between Prony and WKB frequencies. This sets the expected numerical uncertainty well below the main physical trends.

The main result is a clear indication of quasi-resonances, i.e. modes that become arbitrarily long-lived at critical values of $\mu$. This tendency appears for different dilaton couplings. In particular, for $Q=0.7$ and $\mu=0.5$, the damping rate is already very small both for $a=0$ ($\Gamma=0.012075$) and especially for $a=1$ ($\Gamma=0.004618$); in the $\ell=0$ data, near-resonant frequencies are also present, e.g. $\omega=0.202657-0.005137 i$ for $(a,Q,\mu)=(0,0.3,0.25)$ and $\omega=0.202523-0.005226 i$ for $(1,0.3,0.25)$. Very close to the real axis, WKB is not expected to resolve the mode with high precision, but the systematic decrease of $\Gamma(\mu)$ and its extrapolated approach to zero give an unambiguous signal of this regime.

Importantly, the effect of the dilaton coupling is much larger than the estimated error. At fixed $(Q,\mu)=(0.7,0.5)$, changing $a$ from $0$ to $1$ reduces $\Gamma$ from $0.012075$ to $0.004618$ (about $62\%$), while the WKB-order mismatch for $\ell=1$ remains at the level of $\lesssim 1\%$. Therefore, the observed shift from the zero-dilaton case and the emergence of long-lived modes are robust physical effects rather than numerical artifacts.

A dedicated near-axis frequency-domain treatment would be a natural next step in determining the critical $\mu$ values with higher precision. For this purpose, one has to use, for example, the convergent Frobenius method (see, for instance, \cite{Leaver:1985ax,Leaver:1986gd,Nollert:1993zz,Kanti:2006ua,Dias:2022oqm,Pan:2006af,Saka:2025xxl,Lutfuoglu:2025bsf,Stuchlik:2025ezz}). However, for this purpose, the coefficient in the second order wave equation must be reduced to the rational form using either Taylor expansion or parameterization of the metric, as was done, for instance, in \cite{Konoplya:2023ahd,Konoplya:2023aph, Konoplya:2020hyk}. Our work could also be extended to the case of rotating dilatonic black hole given by Kerr-Sen solution \cite{Sen:1992ua,Hioki:2009na,Siahaan:2015ljs}

\begin{acknowledgments}
The author would like to thank Roman Konoplya for encouraging discussions.
\end{acknowledgments}

\bibliography{DilatonMassive-bh}

\end{document}